\documentclass[submission,copyright,creativecommons]{eptcs}
\providecommand{\event}{MeTRiD 2018} 
\usepackage{breakurl}             
\usepackage{underscore}           

\usepackage{amssymb}
\usepackage{graphicx}
\graphicspath{{./eps/}}
\DeclareGraphicsExtensions{.eps}
\usepackage{subfigure}
\usepackage{wrapfig}
\usepackage{amsmath,amsthm}
\usepackage{extarrows}
\usepackage{uppaal}
\usepackage{multirow}
\usepackage{multicol}
\newtheorem{theorem}{Theorem}

\newtheorem{definition}{Definition}

\title{A Compositional Approach for Schedulability Analysis of Distributed Avionics Systems}
\author{Pujie Han \qquad Zhengjun Zhai
\institute{School of Computer Science and Engineering\\
Northwestern Polytechnical University\\
Xi'an, China}
\email{\{hanpujie,zhaizjun\}@mail.nwpu.edu.cn}
\and
Brian Nielsen \qquad Ulrik Nyman
\institute{Department of Computer Science\\
Aalborg University\\
Aalborg, Denmark}
\email{\quad \{bnielsen,ulrik\}@cs.aau.dk}
}
\def\titlerunning{A Compositional Approach for Schedulability Analysis of Distributed Avionics Systems}
\def\authorrunning{P. Han, Z. Zhai B. Nielsen \& U. Nyman}

\begin{document}

\maketitle

\begin{abstract}
This work presents a compositional approach for schedulability analysis of Distributed Integrated Modular Avionics (DIMA) systems that consist of spatially distributed ARINC-653 modules connected by a unified AFDX network. We model a DIMA system as a set of stopwatch automata in \uppaal to verify its schedulability by model checking. However, direct model checking is infeasible due to the large state space. Therefore, we introduce the compositional analysis that checks each partition including its communication environment individually. Based on a notion of message interfaces, a number of message sender automata are built to model the environment for a partition. We define a timed selection simulation relation, which supports the construction of composite message interfaces. By using assume-guarantee reasoning, we ensure that each task meets the deadline and that communication constraints are also fulfilled globally. The approach is applied to the analysis of a concrete DIMA system.
\end{abstract}

\section{Introduction}\label{sec:intro}

The architecture of Distributed Integrated Modular Avionics (DIMA) has been successfully applied to the aviation industry. A DIMA system installs standardized computer modules in spatially distributed locations\cite{wang2013research} that are connected by a unified bus system\cite{annighofer2014systems} such as an AFDX network. Avionics applications residing on the modules run in ARINC-653\cite{arinc653} compliant operating systems. The generic distributed structure of DIMA significantly improves performance and availability as well as reduces development and maintenance costs, while it also dramatically increases the complexity of schedulability analysis. A schedulable DIMA system should fulfil not only the temporal requirements of real-time tasks in each ARINC-653 module but also communication constraints among the distributed nodes. As a result, the system integrators need to consider both computation and communication when analyzing the schedulability of DIMA architecture.

Currently, model checking approaches have been increasingly developed in the schedulability analysis of complex real-time systems. However, we found no studies that analyzed the schedulability of distributed avionics systems as a whole including the network by model checking. The related research isolates computation modules from their underlying network, thereby considering these nodes as independent hierarchical scheduling systems or investigating the network in isolation, which possibly leads to pessimistic results. There have been works using model-checking to analyze the temporal behavior of individual avionics modules in various formal models such as Coloured Petri Nets (CPN)\cite{dodd2006coloured}, preemptive Time Petri Nets (pTPN)\cite{carnevali2011formal}, Timed Automata (TA)\cite{amnell2003times}, and StopWatch Automata (SWA)\cite{mikuvcionis2010schedulability,cicirelli2012development}, and verify schedulability properties via state space exploration. Unfortunately, when being applied to concrete avionics systems, all of them suffer from an inevitable problem of state space explosion. For hierarchical scheduling systems, some studies\cite{carnevali2013compositional,sun2014component,boudjadar2014compositional} exploit the inherent temporal isolation of ARINC-653 partitions\cite{arinc653} and analyze each partition separately, but they ignore the behavior of the underlying network or the interactions among partitions. Thus these methods are not applicable to DIMA environments in which multiple distributed ARINC-653 partitions communicate through a shared network to perform an avionics function together.

In this paper, we present a compositional approach for schedulability analysis of DIMA systems that are modeled as \uppaal SWA, i.e. the TA extended with stopwatches. Compared with the clocks in TA, stopwatches can be blocked and resumed at any location and thus are effective in modeling task preemption. We decompose the system in such a way that we can check each ARINC-653 partition \emph{including} a model of its communication environment individually and then assemble the local results together to derive conclusions about the schedulability of an entire system. Thereby, we verify a number of smaller, simpler, abstract systems rather than directly verifying a larger, more complex, concrete system including the details about all the partitions and the network. The main contributions of this paper are summarized as follows:
\begin{itemize}
  \item \emph{A compositional approach} performs assume-guarantee reasoning\cite{grumberg1994model} to reduce the complexity of symbolic model-checking in the schedulability analysis of DIMA systems.
  \item \emph{An abstraction relation}, timed selection simulation relation, allows users to create a set of abstract models that collectively describe the external behavior of a concrete model, thereby simplifying the abstraction in assume-guarantee reasoning.
  \item A notion of \emph{message interfaces} decouples the communication dependencies between partitions. By composing any partition with its related message interfaces and verifying safety properties of the composition, we can conclude that these properties are still preserved at the global level.
\end{itemize}

The rest of the paper is organized as follows. Section \ref{sec:preliminaries} gives the necessary formal notions. The \uppaal modeling of DIMA systems is presented in section \ref{sec:avionics}. Section \ref{sec:tss} gives the concept of timed selection simulation and its properties. In section \ref{sec:comp}, we detail the compositional analysis approach. Section \ref{sec:experiment} shows an experiment on a concrete DIMA system, and section \ref{sec:conclusion} finally concludes.

\section{Preliminaries}\label{sec:preliminaries}

In this section, we present formal definitions including SWA with an input/output extension and its semantic object Timed I/O Transition Systems(TIOTSs)\cite{david2010timed}.

Suppose that $C$ is a finite set of clocks and $V$ is a finite set of integer variables. A \emph{valuation} $u(x)$ with $x\in C\cup V$ denotes a mapping from $C$ to $\mathbf{R}_{\ge0}$ and from $V$ to $\mathbf{N}$. Let $\mathit{LC}(C,V)$ be the set of linear constraints. A \emph{guard} $g\in\mathit{LC}(C,V)$ is a linear constraint which is defined as a finite conjunction of atomic formulae in the form of $c\sim n$, $c-c^\prime\sim n$ or $v\sim n$ with $c,c^\prime\in C,v\in V,n\in\mathbf{N}$, and $\sim\in\{>,<,=\}$. Given any valuation $u$, we change the values of clocks and integer variables using an \emph{update} operation $r(u)\in2^R$ in the form of $c=0$ or $v=n$ where $c\in C,v\in V$ and $n\in\mathbf{N}$, and $R$ is the set of all possible update operations. In addition, we define an \emph{action} set $\Sigma$. All the actions can be subsumed under two sets of unicast actions $\Sigma^u$ and broadcast actions $\Sigma^b$. By contrast, $\tau\notin\Sigma$ denotes an internal action and $\Sigma^{\tau}=\Sigma\cup\{\tau\}$.

\begin{definition}[Stopwatch Automaton\cite{cassez2000impressive}]
A stopwatch automaton is a tuple $\langle Loc,l_0,C,V,E,\Sigma,Inv,drv\rangle$ where $Loc$ is a finite set of locations, $l_0\in Loc$ is the initial location, $C$ is a finite set of clocks, $V$ is a finite set of integer variables, $E\subseteq Loc\times\mathit{LC}(C,V)\times\Sigma^{\tau}\times2^R\times Loc$ is a set of edges, $\Sigma=I\oplus O$ is a finite set of actions divided into inputs($I$) and outputs($O$), $Inv$ is a mapping $Loc\to\mathit{LC}(C,V)$, and $drv$ is a mapping $Loc\times C\to\{0,1\}$.
\end{definition}

From a syntactic viewpoint, SWA belongs to the class of TA extended with $drv$, which can prevent part of the clocks from changing in specified locations semantically. We now shift the focus to the semantic object TIOTS of SWA.

In a TIOTS, there are two types of transitions: delay and action transitions. We use the set $D=\{\epsilon(d)|d\in\mathbf{R}_{\ge0}\}$ to denote the delay, and refer to the 0-delay $\epsilon(0)$ as $\mathbf{0}$.

\begin{definition}[Timed I/O Transition System]\label{def:tiots}
A timed I/O transition system is a tuple $\mathcal{T}=\langle S,s_0,\Sigma,\to\rangle$ where $S$ is an infinite set of states, $s_0$ is the initial state, $\Sigma=I\oplus O$ is a finite set of actions divided into inputs($I$) and outputs($O$), $I\cap O\subseteq\Sigma^u$, and $\to\subseteq S\times\Sigma^{\tau}\cup D\times S$ is a transition relation. $s\xrightarrow{a} s^\prime$ represents $(s,a,s^\prime)\in \to$, which has the properties of time determinism, time reflexivity, and time additivity\cite{david2010timed}.
\end{definition}

For any SWA, a state is defined as a pair $\langle l,u\rangle$ where $l$ is a location and $u$ is a valuation over clocks and integer variables. On the basis of TIOTSs, the operational semantics of SWA is defined as follows.

\begin{definition}
The operational semantics of a stopwatch automaton $A=\langle Loc,l_0,C,V,E,\Sigma,Inv,drv\rangle$ is a timed I/O transition system $\mathcal{T}^A=\langle S,s_0,\Sigma,\to\rangle$ where $S$ is the set of states of $A$, $s_0=\langle l_0,u_0\rangle$ is the initial state of $A$, $\Sigma$ is the same set of actions as $A$, and $\to$ is the transition relation defined by
\begin{itemize}
\item $\langle l,u\rangle\xrightarrow{a}\langle l^\prime,u^\prime\rangle$ iff $\exists\langle l,g,a,r,l^\prime\rangle\in E\ $ $(u\models g\ \land\ u^\prime=r(u)\ \land\ u^\prime\models Inv(l^\prime))$
\item $\langle l,u\rangle\xrightarrow{\epsilon(d)}\langle l^\prime,u^\prime\rangle$ iff $l=l^\prime\land(\forall v\in V\ $ $u^\prime(v)=u(v))\ \land\ (\forall c\in C\ (drv(l,c)=0\Rightarrow u^\prime(c)=u(c)))\ \land\ (\forall c\in C\ (drv(l,c)=1\Rightarrow u^\prime(c)=u(c)+d))\ \land\ u^\prime\models Inv(l^\prime)$.
\end{itemize}
\end{definition}

For any transition $s\xrightarrow{a} s^\prime$, two symbols $a?$ and $a!$ denote the action $a$ belonging to input $I$ and output $O$ respectively. Given $a\in\Sigma$, $s\xrightarrow{a}$ iff $\exists s^\prime\in S$, s.t. $s\xrightarrow{a}s^\prime$. $\xrightarrow{\tau}^{\ast}$ or $\xLongrightarrow{\mathbf{0}}$ denotes the reflexive and transitive closure of $\xrightarrow{\tau}$. $s\xLongrightarrow{\epsilon{(d)}}s^\prime$ iff $s\xrightarrow{\epsilon{(d)}}s^\prime$, or $\exists s_1,s_2,\dots,s_n\in S$, s.t. $s\xrightarrow{\alpha_0}s_1\xrightarrow{\alpha_1}s_2\xrightarrow{\alpha_2}\cdots\xrightarrow{\alpha_{n-1}}s_n\xrightarrow{\alpha_n}s^\prime$ and $\forall i\in\{0,\dots,n\}$, s.t. $\alpha_i=\tau$ or $\alpha_i\in D$ and $d=\sum\{d_i|\alpha_i=\epsilon(d_i)\}$.

The definition of parallel composition $\|$ of TIOTSs is similar to that in \cite{david2010timed}. Given two TIOTSs $\mathcal{T}_i=\langle S_i,s_{i,0},\Sigma_i,\to_i\rangle,i\in\{1,2\}$, they are \emph{compatible} iff they satisfy the following conditions:
\begin{itemize}
\item \emph{(Unique output)} $O_1\cap O_2=\varnothing$.
\item \emph{(Deterministic-pair unicast)} $I_1\cap I_2\cap\Sigma^u=\varnothing$.
\end{itemize}
Note that broadcast actions in the composition of TIOTSs are \emph{input-enabled}: $\forall s\in S_i\ \forall a\in I_i\cap\Sigma^b\ s\xrightarrow{a}$.

\begin{definition}[Parallel Composition]
Suppose two timed I/O transition systems $\mathcal{T}_1=\langle S_1,s_{1,0},\Sigma_1,\to_1\rangle$ and $\mathcal{T}_2=\langle S_2,s_{2,0},\Sigma_2,\to_2\rangle$ are compatible. The parallel composition $\mathcal{T}_1||\mathcal{T}_2$ is the timed I/O transition system $\langle S,s_0,\Sigma,\to\rangle$ where $S=S_1\times S_2$, $s_0=\langle s_{1,0},s_{2,0}\rangle$, $\Sigma=I_{1||2}\oplus O_{1||2}$, $I_{1||2}=(I_1\setminus(O_2\cap\Sigma^b))\cup (I_2\setminus (O_1\cap\Sigma^b))$, $O_{1||2}=O_1\cup O_2$, and $\to$ is the largest relation generated by the following rules:
\begin{itemize}
\item \textit{INDEP-L:}$\begin{aligned}[t]\quad\frac{s_1\xrightarrow{a}s_1^\prime\quad a\in\{\tau\}\cup\Sigma_1\setminus\Sigma_2}{\langle s_1,s_2\rangle\xrightarrow{a}\langle s_1^\prime,s_2\rangle}\quad\quad\textit{INDEP-R:}\quad\frac{s_2\xrightarrow{a}s_2^\prime\quad a\in\{\tau\}\cup\Sigma_2\setminus\Sigma_1}{\langle s_1,s_2\rangle\xrightarrow{a}\langle s_1,s_2^\prime\rangle}\end{aligned}$       \\
\item \textit{DELAY:}$\begin{aligned}[t]\quad\frac{s_1\xrightarrow{\epsilon(d)}s_1^\prime\quad s_2\xrightarrow{\epsilon(d)}s_2^\prime\quad d\in\mathbf{R}_{\ge0}}{\langle s_1,s_2\rangle\xrightarrow{\epsilon(d)}\langle s_1^\prime,s_2^\prime\rangle}\end{aligned}$              \\
\item \textit{SYNC-IN:}$\begin{aligned}[t]\quad\frac{s_1\xrightarrow{a}s_1^\prime\quad s_2\xrightarrow{a}s_2^\prime\quad a\in I_{1||2}}{\langle s_1,s_2\rangle\xrightarrow{a}\langle s_1^\prime,s_2^\prime\rangle}\end{aligned}$       \\
\item \textit{SYNC-BIO:}$\begin{aligned}[t]\quad\frac{s_1\xrightarrow{a}s_1^\prime\quad s_2\xrightarrow{a}s_2^\prime\quad a\in(I_1\cap O_2)\cup(O_1\cap I_2)\cap\Sigma^b}{\langle s_1,s_2\rangle\xrightarrow{a}\langle s_1^\prime,s_2^\prime\rangle}\end{aligned}$       \\
\item \textit{SYNC-UIO:}$\begin{aligned}[t]\quad\frac{s_1\xrightarrow{a}s_1^\prime\quad s_2\xrightarrow{a}s_2^\prime\quad a\in I_{1||2}\cap O_{1||2}}{\langle s_1,s_2\rangle\xrightarrow{\tau}\langle s_1^\prime,s_2^\prime\rangle}.\end{aligned}$
\end{itemize}
\end{definition}

We use $\Omega$ to denote the set of TA and SWA in our modeling framework. For any $A,B\in\Omega$, we define the composite model $C=A\|B$ iff their TIOTSs satisfy $\mathcal{T}^C=\mathcal{T}^A||\mathcal{T}^B$.

\section{Avionics System Modeling}\label{sec:avionics}

\begin{wrapfigure}{r}{0.4\textwidth}
\vspace{-20pt}
\centering
\includegraphics[width=2.1in]{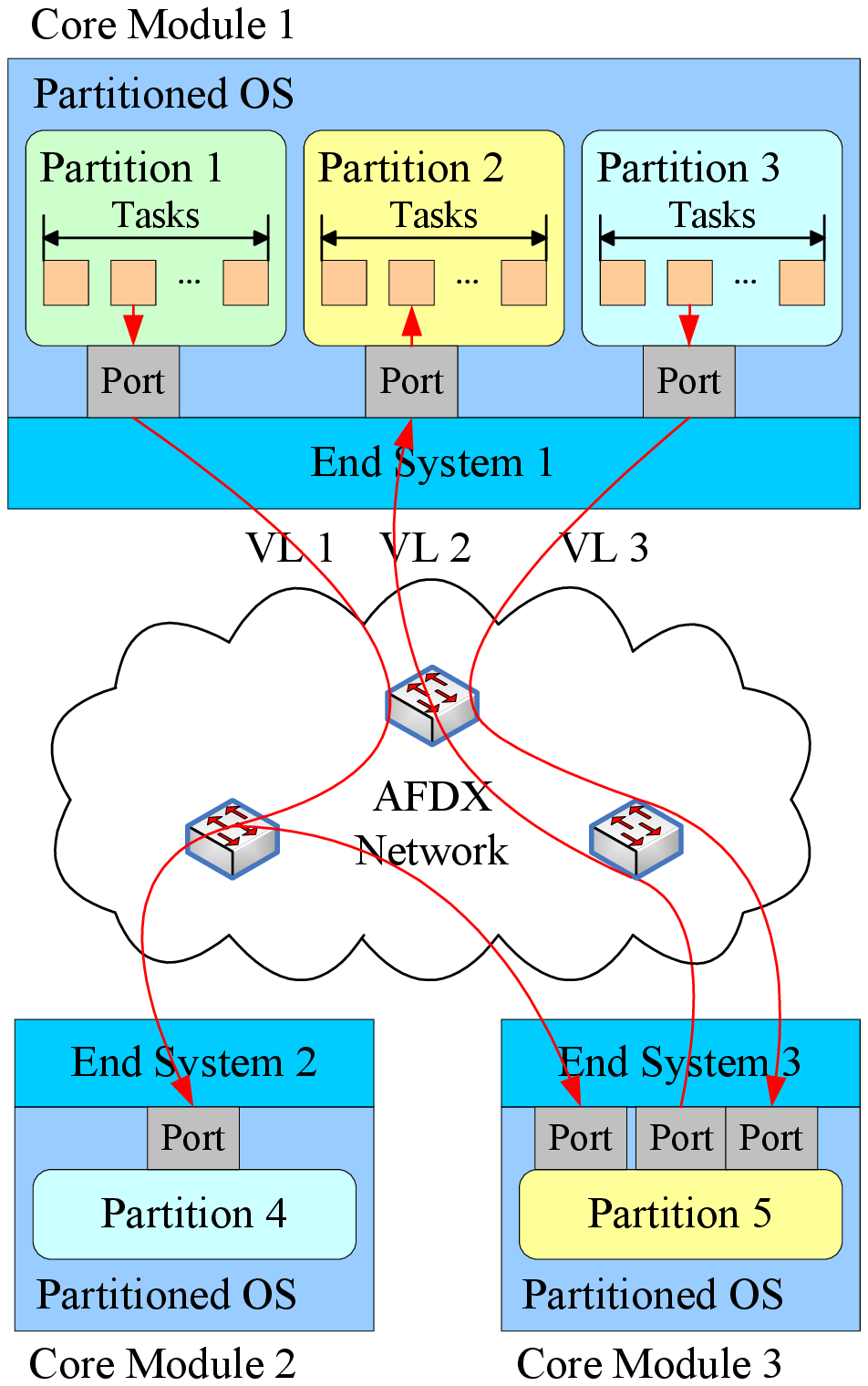}
\caption{An Example of DIMA systems}
\label{fig:afdx}
\vspace{-10pt}
\end{wrapfigure}

We focus on a generic DIMA architecture including a set of ARINC-653 modules connected by an AFDX network, as shown in Fig.\ref{fig:afdx}. There is a three-layer structure in the DIMA system that consists of scheduling, task, and communication layers.

\begin{figure}[!t]
\centering
\includegraphics[width=3.5in]{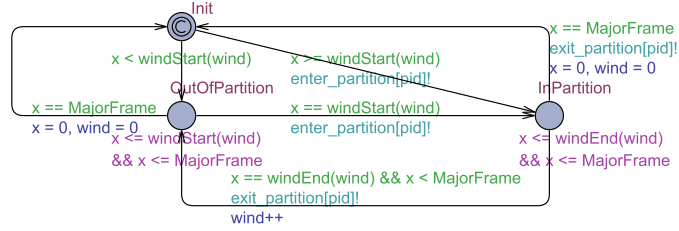}
\caption{The \uppaal Template of an ARINC-653 Partition Scheduler}
\label{fig:ps}
\end{figure}

The \emph{scheduling layer} is defined as the scheduling facilities for generic computation resources of a DIMA system, where standardized computer modules execute concurrent application tasks in partitioned operating systems. In this operating system, partitions are scheduled by a Time Division Multiplexing (TDM) scheduler and each partition also has its local scheduling policy, preemptive Fixed Priority (FP), to manage the internal tasks\cite{arinc653}. The scheduling layer is modeled as two TA templates \uppTemp{PartitionSupply} and \uppTemp{TaskScheduler} in \uppaal\footnote{Models available at \url{http://eptcs.web.cse.unsw.edu.au/paper.cgi?MARSVPT2018:2}}. The \uppTemp{PartitionSupply} depicted in Fig.\ref{fig:ps} provides the service of TDM partitioning for a particular partition \uppVar{pid}. The \uppTemp{TaskScheduler} implementing FP scheduling allocates processor time to the task layer only when the partition is active.

The \emph{task layer} contains all the application tasks executing avionics functions. A task is regarded as the smallest scheduling unit, each of which runs concurrently with other tasks in the same partition. The execution of a task is modelled as a sequence of commands that are either computing for a duration, locking/unlocking a resource, or sending/receiving a message. We consider two task types: \emph{periodic tasks} and \emph{sporadic tasks}. A periodic task has a fixed release period, while a sporadic task is characterized by a minimum separation between consecutive jobs. The task layer is instantiated from two SWA templates \uppTemp{PeriodicTask} and \uppTemp{SporadicTask} in \uppaal. Since the tasks in a partition are scheduled by a task scheduler, we use a set of binary channels as scheduling actions to communicate between task models and \uppTemp{TaskScheduler}.

The \emph{communication layer} carries out inter-partition communication over a common AFDX network. The AFDX protocol stack realized by an End System(ES) interfaces with the task layer through ARINC-653 ports. Based on the AFDX protocol structure, the communication layer is further divided into UDP/IP layer and Virtual Link layer, where a Virtual Link (VL) ensures an upper bound on end-to-end delay. In \uppaal, the UDP/IP layer is divided into two TA templates \uppTemp{IPTx} and \uppTemp{IPRx}, which calculate the latency of the UDP/IP layer in a transmitting ES and a receiving ES respectively. Similarly, two TA templates \uppTemp{VLinkTx} and \uppTemp{VLinkRx} model the delay of a VL in opposite directions.

From a global view of the system, its schedulability is also affected by the communication layer. According to the ARINC-653 standard\cite{arinc653}, there are two types of ARINC-653 ports, sampling ports and queuing ports. A sampling port can accommodate at most a single message that remains until it is overwritten by a new message. A refresh period is defined for each sampling port. This attribute provides a specified arrival rate of messages, regardless of the rate of receiving requests from tasks. In contrast, a queuing port is allowed to buffer multiple messages in a message queue with a fixed capacity. However, the operating system is not responsible for handling overflow from the message queue.

In this paper, we verify the following three typical schedulability properties:
\begin{itemize}
  \item All the tasks meet their deadlines in each partition.
  \item The refresh period of any sampling port is guaranteed.
  \item The overflow from any queuing ports must be avoided.
\end{itemize}

The schedulability of an avionics system is described and verified as a safety property of the above TA/SWA models. We add a set $Err$ of \emph{error locations} to the templates. Once schedulability is violated, the related model will lead itself to one of the error locations immediately. Thus, the schedulability is replaced with this safety property $\varphi$:
\begin{equation}\label{eq:safetylogic}
\textrm{\uppPropAG{}}\ \lnot(\bigvee\nolimits_{loc\in Err} loc),
\end{equation}
which belongs to a simplified subset of TCTL used in \uppaal.

However, since the verification algorithm inside \uppaal for SWA introduces a slight over-\-appr\-oxim\-ation\cite{cassez2000impressive}\footnote{Exact reachability for SWA with more than 3 stopwatches is known to be undecidable\cite{cassez2000impressive}.}, \uppaal may sometimes give the verification result ``Maybe satisfied'' or ``May not be satisfied''. To further refine the result in this case we manually analyse the possible counter example using \uppaal{}'s concrete simulator to determine if the system is unschedulable. Alternatively, the statistical model-checking (SMC) engine could be invoked to attempt an automatic falsification. In our experiences, the result only appears when the system is on the very borderline of being schedulable.

\section{Timed Selection Simulation}\label{sec:tss}

We propose a notion of timed selection simulation relation to support assume-guarantee reasoning. Compared with some other abstraction relations like timed simulation\cite{jensen2000scaling} and timed ready simulation\cite{jensen1999abstraction}, timed selection simulation only abstracts a selected subset of actions from the concrete model. Applying timed selection simulation to the abstraction of a concrete system, one can pay attention to part of the system, individually model the behavior of each component, and thereby obtain a composite abstract model rather than a monolithic one.

Considering the semantic object $\mathcal{T}^A$ of an automaton $A\in\Omega$, we denote the \emph{error states} of $\mathcal{T}^A$ by the set $\mathcal{E}=\{\langle l,u\rangle|l\in Err\}$ where $Err$ is the error-location set of $A$. Thus, for any TIOTS $\mathcal{T}=\langle S,s_0,\Sigma,\to\rangle$, its error states are defined as a set $\mathcal{E}\subseteq S$, and the following function $g:S\to\{\mathit{true},\mathit{false}\}$ indicates whether a state $s\in S$ has violated schedulability properties:
\begin{equation}
g(s)=\left\{\begin{array}{ll}
    \mathit{true} &\ \textrm{if $s\in\mathcal{E}$}\\
    \mathit{false} &\ \textrm{if $s\not\in\mathcal{E}$}.
    \end{array}\right.
\end{equation}
Given two compatible TIOTSs $\mathcal{T}_i,i\in\{1,2\}$ with the error-state set $\mathcal{E}_i$, their composition $\mathcal{T}_1\|\mathcal{T}_2$ has the error-state set $\mathcal{E}_{\mathcal{T}_1\|\mathcal{T}_2}=\{\langle s_1,s_2\rangle|s_1\in\mathcal{E}_1\lor s_2\in\mathcal{E}_2\}$ and the function $g(\langle s_1,s_2\rangle)=g(s_1)\lor g(s_2)$.

Based on the function $g(s)$, the formal definition of timed selection simulation is given as follows.

\begin{definition}[Timed Selection Simulation]\label{def:tss}
Let $\mathcal{T}_1=\langle S_1,s_{1,0},\Sigma_1,\to_1\rangle$ and $\mathcal{T}_2=\langle S_2,$ $s_{2,0},\Sigma_2,\to_2\rangle$ be two timed I/O transition systems with $\Sigma_2\subseteq\Sigma_1$. Let R be a relation from $S_1$ to $S_2$. We call R a timed selection simulation from $\mathcal{T}_1$ to $\mathcal{T}_2$, written $\mathcal{T}_1\preceq \mathcal{T}_2$ via $R$, provided $(s_{1,0},s_{2,0})\in R$ and for all $(s_1,s_2)\in R$, $g(s_1)=g(s_2)$ and
\begin{enumerate}
\item if $s_1\xrightarrow{a?}s_1^\prime$ for some $s_1^\prime\in S_1$, $a\in\Sigma_2$, then $\exists s_2^\prime\in S_2$ such that $s_2\xLongrightarrow{a?}s_2^\prime$ and $(s_1^\prime,s_2^\prime)\in R$
\item if $s_1\xrightarrow{a!}s_1^\prime$ for some $s_1^\prime\in S_1$, $a\in\Sigma_2$, then $\exists s_2^\prime\in S_2$ such that $s_2\xLongrightarrow{a!}s_2^\prime$ and $(s_1^\prime,s_2^\prime)\in R$
\item if $s_1\xrightarrow{a}s_1^\prime$ for some $s_1^\prime\in S_1$, $a\in(\Sigma_1\setminus\Sigma_2)\cup\{\tau\}$, then $\exists s_2^\prime\in S_2$ such that $s_2\xLongrightarrow{\mathbf{0}}s_2^\prime$ and $(s_1^\prime,s_2^\prime)\in R$
\item if $s_1\xrightarrow{\epsilon(d)}s_1^\prime$ for some $s_1^\prime\in S_1$, $d>0$, then $\exists s_2^\prime\in S_2$ such that $s_2\xLongrightarrow{\epsilon(d)}s_2^\prime$ and $(s_1^\prime,s_2^\prime)\in R.$
\end{enumerate}
\end{definition}

\begin{definition}
Let $A_i, i\in\{1,2\}$ be stopwatch automata. We say that $A_1\preceq A_2$, if and only if their corresponding timed I/O transition systems $\mathcal{T}_i$ satisfy $\mathcal{T}_1\preceq\mathcal{T}_2$.
\end{definition}

We now give some necessary properties of timed selection simulation.

\begin{theorem}\label{thm:preorder}
Timed selection simulation $\preceq$ is a preorder.
\end{theorem}

For any automaton $A\in\Omega$, by construction, the reachability of its error locations is equivalent to that of the error states in the corresponding TIOTS $\mathcal{T}^A$. Hence the following theorem shows that timed selection simulation can preserve the satisfaction of the safety properties in the form of Eq.(\ref{eq:safetylogic}).

\begin{theorem}[Property preservation]\label{thm:preservation}
Let $\mathcal{T}_i, i\in\{$ $1,2\}$ be timed I/O transition systems and $\mathcal{E}_i$ be the set of error states of $\mathcal{T}_i$. Given a safety property $\varphi:\lnot reach(\mathcal{E}_i)$ that any error states are not reachable, if $\mathcal{T}_1\preceq\mathcal{T}_2$ and $\mathcal{T}_2\models\varphi$, then $\mathcal{T}_1\models\varphi$.
\end{theorem}

\begin{theorem}[Abstraction compositionality]\label{thm:abscomp}
Let $\mathcal{T}_i,i\in\{1,2,3\}$ be timed I/O transition systems. If $\mathcal{T}_1\preceq\mathcal{T}_2$, $\mathcal{T}_1\preceq\mathcal{T}_3$, and $\mathcal{T}_2$ and $\mathcal{T}_3$ are compatible, then $\mathcal{T}_1\preceq\mathcal{T}_2\|\mathcal{T}_3$.
\end{theorem}

\begin{theorem}[Compositionality]\label{thm:comp}
Let $\mathcal{T}_i=\langle S_i,s_{i,0},$ $\Sigma_i,\to_i\rangle$, $i\in\{1,2,3,4\}$ be timed I/O transition systems. Suppose $\mathcal{T}_1\|\mathcal{T}_3$ and $\mathcal{T}_2\|\mathcal{T}_4$ are the parallel compositions of compatible timed I/O transition systems. If $(1)\ \mathcal{T}_1\preceq\mathcal{T}_2, \mathcal{T}_3\preceq\mathcal{T}_4$, and $(2)\ O_1\cap I_4\subseteq\Sigma_2\subseteq\Sigma^b, I_2\cap O_3\subseteq\Sigma_4\subseteq\Sigma^b$, then $\mathcal{T}_1\|\mathcal{T}_3\preceq\mathcal{T}_2\|\mathcal{T}_4$.
\end{theorem}

\section{Compositional Analysis}\label{sec:comp}

\begin{figure}[!t]
\centering
\includegraphics[width=.7\textwidth]{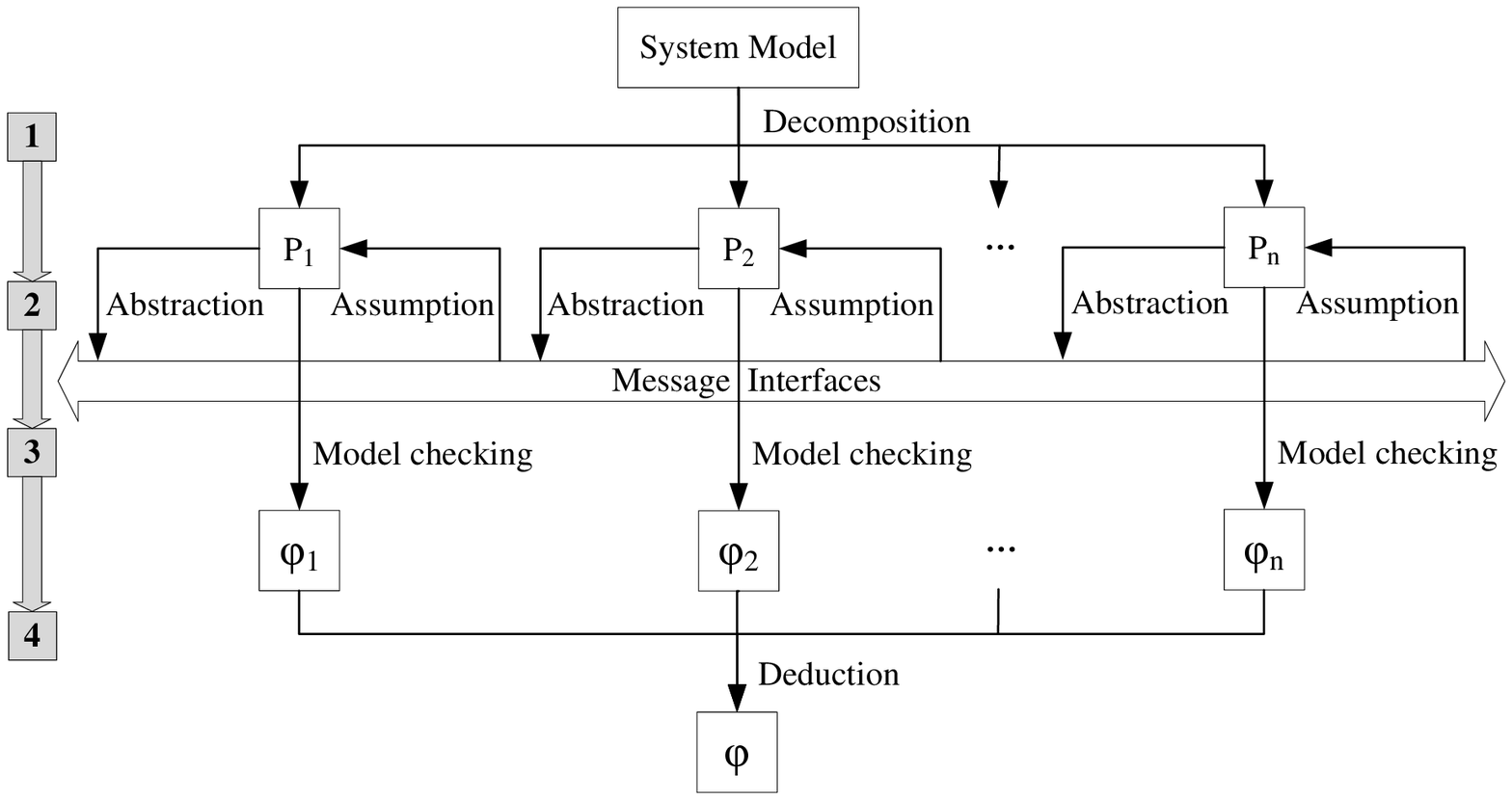}
\caption{Compositional Analysis Procedure}
\label{fig:comp}
\end{figure}

We apply assume-guarantee reasoning to the schedulability analysis, and describe the schedulability goal as a safety property $\varphi$ (Eq.(\ref{eq:safetylogic})). As shown in Fig.\ref{fig:comp}, our compositional analysis is comprised of the following four steps:
\begin{enumerate}
\item \emph{Decomposition:} The system is first decomposed into a set of communicating partitions modeled by TA and SWA. The global property $\varphi$ is also divided into several local properties, each of which belongs to one partition.
\item \emph{Construction of message interfaces:} We define message interfaces as the assumption and abstraction of the communication environment for each partition. In general, the templates of message interfaces should be built manually by the engineers.
\item \emph{Model checking:} The local properties under the assumptions and the abstraction relations are verified by model checking.
\item \emph{Deduction:} From the assume-guarantee rules, we finally derive the global property by combining all the local results.
\end{enumerate}

The procedure can be performed automatically except for the first construction of message interfaces. We assume that a task never blocks while communicating with other partitions, which is commonly used in avionics systems\cite{easwaran2009compositional,carnevali2013compositional}. Otherwise a loop of communication dependency will cause circular reasoning, because the assumptions of a partition might be based on its own state recursively.

\subsection{Decomposition}

Assume that there are $n$ constituent partitions in a system. Let $P_i,i\in\{1,2,\dots,n\}$ be the SWA composite model of a partition. Let $Err_i$ be the error-location set of $P_i$. The safety property $\varphi_i$: $\textrm{\uppPropAG{}}\ \lnot(\bigvee_{loc\in Err_i} loc)$ denotes the schedulability of $P_i$. The global property $\varphi$ is therefore written as $\varphi_1\land\varphi_2\land\cdots\land\varphi_n$, and the goal of our schedulability analysis is expressed as the verification problem:
\begin{equation}
P_1\|P_2\|\cdots\|P_n\models\varphi
\end{equation}
that can be further divided into $n$ satisfaction relations:
\begin{equation}\label{eq:subqs}
P_1\|P_2\|\cdots\|P_n\models\varphi_i,\ i\in\{1,2,\ldots,n\}.
\end{equation}

Since the error-location set $Err_i$ is only allowed to be manipulated by $P_i$, we check each partition model $P_i$ independently for the corresponding \emph{local property} $\varphi_i$ instead of the original verification problem with a large and complex system. However, the communication environment of $P_i$, which denotes the behavior that $P_i$ receives messages from other partitions, may affect the satisfaction of the schedulability property $\varphi_i$. Hence when performing the verification for partition $P_i$, one needs to give the \emph{assumptions} of its communication environment and verifies the local property $\varphi_i$ under these assumptions.

\subsection{Construction of message interfaces}

A set of TA models is created to describe the message-sending behavior of a partition. Each of the TA is called a \emph{message interface} of this partition and associated with a particular message type. Suppose there are a number of messages sent from partition $P_j$ to another partition $P_i$ and their corresponding message interfaces make up a composite TA model $A_{i,j}$. When we analyze $P_i$ in the compositional way, it should be safe for $A_{i,j}$ to replace $P_j$. Hence, we say that a message interface of $P_j$ is an \emph{abstraction} of $P_j$.

Our abstraction of the message delivery between a partition and its underlying network is modelled using broadcast synchronization. A broadcast action represents a specific message types. Let $\Sigma_i=I_i\oplus O_i$ be the action set of a composite model for any partition $P_i$. An action $a_k\in I_i\cap\Sigma^b$(resp. $a_k\in O_i\cap\Sigma^b$) denotes that $P_i$ receives(resp. sends) messages with the type $msg_k$ from(resp. to) other partition(s). The symbol $j\triangleright i$ represents the condition that there exists a partition $P_j$ sending messages to $P_i$ via an action set $O_{j\rightharpoonup i}\subseteq I_i\cap O_j$.

\begin{definition}[Message Interface]
Let $O_i$ be the output action set of a stopwatch automaton $P_i\in\Omega$. For any output action $a_k\in O_i\cap\Sigma^b$, the timed automaton $A^k_i$ with an action set $\Sigma^k_i=O^k_i=\{a_k\}$ is a message interface of $P_i$ if and only if there exists a timed selection simulation relation $\preceq$ on $\Omega$ such that
\begin{equation}\label{eq:absmsg}
P_i\preceq A^k_i.
\end{equation}
\end{definition}

\begin{figure}[!t]
\centering
\includegraphics[width=.5\textwidth]{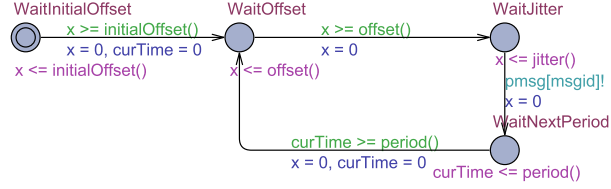}
\caption{An Example of a Message Interface}
\label{fig:mieg}
\end{figure}
We build the templates of message interfaces in accordance with the characteristics of message-sending actions. In practice, the structure of an interface can be designed straightforwardly from the task specification. The template in Fig.\ref{fig:mieg} shows a message interface that sends messages periodically via the action array \uppSync{pmsg}. Then we make an automatized binary search for the interface's parameters such as \uppConst{offset} in the template and meanwhile check the satisfaction of timed selection simulation relation.

The message interfaces can serve as the assumptions of the communication environment of a partition. The composition $A_{i,j}$ of the message interfaces $A^k_j$ for all $a_k\in O_{j\rightharpoonup i}$ provides $P_i$ with a ``complete'' abstraction of $P_j$, which models the behavior of all the output actions from $P_j$ to $P_i$. According to the abstraction compositionality (Theorem \ref{thm:abscomp}) of the preorder $\preceq$, we have
\begin{equation}\label{eq:abs}
P_j\preceq A_{i,j}.
\end{equation}
Considering all the partitions except $P_i$ in the system, we describe the communication environment of $P_i$ as the composite model $\big\|_{j=1,j\neq i}^{n}A_{i,j}$.

\subsection{Model checking}

In the third step, the local property $\varphi_i$ of $P_i$ under assumption $\big\|_{j=1,j\neq i}^{n}A_{i,j}$ can be verified by model checking. We denote these $n$ subproblems by
\begin{equation}\label{eq:mcqs}
P_i\ \|\ (\Big\|_{j=1,j\neq i}^{n}A_{i,j})\models\varphi_i\ \ i\in\{1,2,\dots,n\}.
\end{equation}
Normally, $A_{i,j}$ in Eq.(\ref{eq:mcqs}) has a much smaller model size than its corresponding partition model $P_j$ in Eq.(\ref{eq:subqs}). Thus, the compositional approach allows us to verify a simpler abstract partition model instead of a complex concrete system model including the details about all the partitions.

In addition, we capture the computation time of each task as an interval between a best-case and worst-case execution time. When analyzing the schedulability of a partition, the model-checker explores all scheduling decisions that can be made in such an interval, and hence also examines possible cases of scheduling timing anomalies\cite{Reineke2006A}.

\subsection{Deduction}\label{sec:deduction}

We derive the global property $\varphi$ by combining $n$ local results in the last step. For any schedulable system, each property $\varphi_i$ should be concluded from the satisfaction of Eq.(\ref{eq:mcqs}) under assumptions and all the abstraction relations of Eq.(\ref{eq:abs}). According to the compositionality (Theorem \ref{thm:comp}) and property preservation (Theorem \ref{thm:preservation}) of timed selection simulation, we have the following assume-guarantee rule:
\begin{equation}\label{eq:deductionrule}
\frac{
\begin{split}
\bigwedge\nolimits_{\{j|j\triangleright i\}} P_j\preceq A_{i,j}\\
P_i\ \|\ (\Big\|_{j=1,j\neq i}^{n}A_{i,j})\models\varphi_i
\end{split}
}
{\hspace{8mm}P_1\|P_2\|\cdots\|P_n\models\varphi_i}
\end{equation}
Note that this assume-guarantee rule only provides a sufficient schedulability condition, for abstract message interfaces might slightly over-approximate the external behavior of a partition.

A simplified DIMA system exemplifies the reasoning procedure. In the example, the system model is decomposed into three partitions $P_i,i\in\{1,2,3\}$. We divide the global property $\varphi$ into three local properties $\varphi_i,i\in\{1,2,3\}$. Accordingly, the goal of the verification problem is to check
\begin{equation}\label{eq:expgoal}
P_1\|P_2\|P_3\models\varphi_1\land\varphi_2\land\varphi_3.
\end{equation}
From Eq.(\ref{eq:subqs}), this problem can be replaced with three subproblems:
\begin{equation}\label{eq:expsubqs}
P_1\|P_2\|P_3\models\varphi_i,i\in\{1,2,3\}.
\end{equation}

Without loss of generality, we take the verification of $\varphi_1$ for example to show how the model-checking and deduction are carried out in the following steps.

Assume that $P_2$ sends $P_1$ two types of messages, $msg_1$ and $msg_2$, via two actions $a_1$ and $a_2$ respectively, and $P_3$ sends $P_1$ only a $msg_3$ with action $a_3$. We create one message interface $A^k_j,j\in\{2,3\}$(like Eq.(\ref{eq:absmsg})) for each message type $msg_k(k\in\{1,2,3\})$ received by $P_1$ in the system. The abstraction relations from Eq.(\ref{eq:absmsg}) can be expressed as
\begin{equation}\label{eq:expabs1}
P_2\preceq A^1_2,\ P_2\preceq A^2_2,\ P_3\preceq A^3_3.
\end{equation}
From abstraction compositionality of the preorder $\preceq$, we can obtain
\begin{equation}\label{eq:abscomp}
P_2\preceq A^1_2\|A^2_2,\ P_3\preceq A^3_3.
\end{equation}
Then, from reflexivity and compositionality of the preorder $\preceq$, the composite model of the system satisfies
\begin{equation}\label{eq:applycomp}
P_1\|P_2\|P_3\preceq P_1\|A^1_2\|A^2_2\|A^3_3.
\end{equation}
Note that when we apply the compositionality to checking a partition $P_i$, any output actions sent to $P_i$ will never be removed in abstraction relations (Eq.(\ref{eq:abscomp})), which satisfies the condition (2) of theorem \ref{thm:comp}.

With Eq.(\ref{eq:applycomp}), we have from property preservation of the abstraction relation $\preceq$ that $\mathit{if}$
\begin{equation}\label{eq:expchk1}
P_1\|A^1_2\|A^2_2\|A^3_3\models\varphi_1, ~\mathit{then}
\end{equation}
\begin{equation}\label{eq:expgoal1}
P_1\|P_2\|P_3\models\varphi_1.
\end{equation}

Since Eq.(\ref{eq:expgoal1}) covering all three partitions in the system has a higher complexity than Eq.(\ref{eq:expchk1}), the techniques of model checking can be adopted to verify the simpler problem Eq.(\ref{eq:expchk1}) instead of the original goal Eq.(\ref{eq:expgoal1}). The same steps will be repeated for local properties $\varphi_2$ and $\varphi_3$.

Consequently, we conclude all the local results of (\ref{eq:expsubqs}) according to the reasoning process from Eq.(\ref{eq:expabs1}) to Eq.(\ref{eq:expgoal1}). When we analyze the partition $P_1$ and its communication environment, the local result of Eq.(\ref{eq:expgoal1}) can be deduced from Eq.(\ref{eq:expabs1}) and Eq.(\ref{eq:expchk1}) in the following assume-guarantee rule.
\begin{equation}\label{eq:exprule}
\frac{
\begin{split}
P_2\preceq A^1_2\land P_2\preceq A^2_2\land P_3\preceq A^3_3\\
P_1\|A^1_2\|A^2_2\|A^3_3\models\varphi_1
\end{split}
}
{\hspace{21mm}P_1\|P_2\|P_3\models\varphi_1}
\end{equation}
The local results are then combined to constitute the global result of Eq.(\ref{eq:expgoal}).

\section{Case Study}\label{sec:experiment}

In this section, we applies the compositional approach to an avionics system which combines the workload of \cite{carnevali2013compositional} and the AFDX configuration of \cite{gutierrez2014holistic}. The workload consists of 5 partitions, and further divided into 18 periodic tasks and 4 sporadic tasks. Considering the inter-partition messages in the workload, we assign each message type $\mathit{Msg_i},i=\{1,2,3,4\}$ a separate VL with the same subscript. The messages of $\mathit{Msg_1}$ and $\mathit{Msg_2}$ are handled at the refresh period $50ms$ in sampling ports. $\mathit{Msg_3}$ and $\mathit{Msg_4}$ are configured to operate in queuing ports, each of which can accommodate a maximum of one message.

\begin{figure*}[!h]
\begin{minipage}{.5\textwidth}
\centering
\includegraphics[width=2.9in]{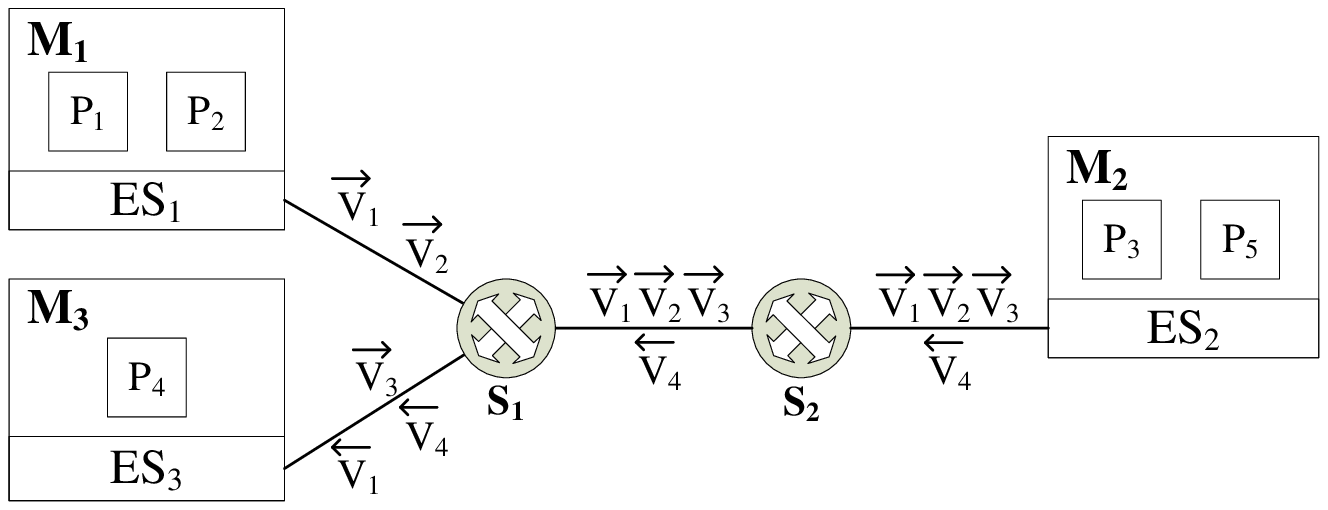}
\end{minipage}%
\begin{minipage}{.5\textwidth}
\centering
\includegraphics[width=2.5in]{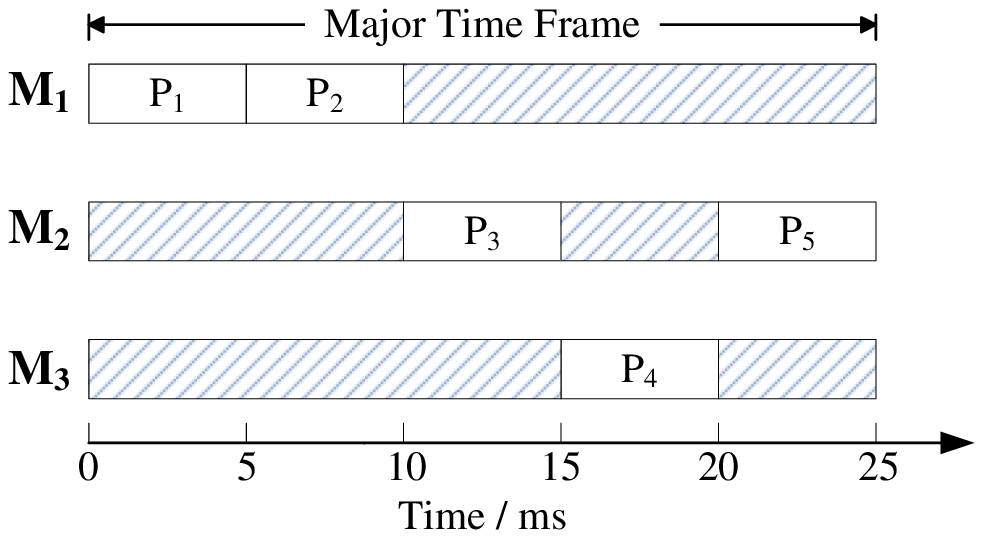}
\end{minipage}%
\caption{The Distributed Avionics Deployment and Partition Schedules (Times in Milliseconds)}
\label{fig:environ}
\label{fig:partsched}
\end{figure*}

As shown in Fig.\ref{fig:environ}, we consider the distributed architecture that comprises 3 ARINC-653 modules connected by an AFDX network. The module $M_1$ accommodates $P_1$ and $P_2$, the module $M_2$ executes $P_3$ and $P_5$, and the partition $P_4$ is allocated to $M_3$. There are 4 VLs $V_1$-$V_4$ connecting 3 ESs across 2 switches $S_1$ and $S_2$ in the AFDX network. The arrows above VLs' names indicate the direction of message flow.

The avionics system equips each of its processor cores with a partition schedule. Assume the modules in the experiment to be single-processor platforms. Fig.\ref{fig:partsched} gives the partition schedules, which fix a common major time frame $T_{mf}$ at $25ms$ and allocate $5ms$ to each partition within every $T_{mf}$. All the partition schedules are enabled at the same initial instant. The scheduling configuration keeps the temporal order of the partitions in \cite{carnevali2013compositional}. Hence the partition schedules contain five disjoint windows $\langle P_1,0,5\rangle$, $\langle P_2,5,5\rangle$, $\langle P_3,10,5\rangle$, $\langle P_4,15,5\rangle$, and $\langle P_5,20,5\rangle$, where the second parameter is the offset from the start of $T_{mf}$ and last the duration.

We analyze the schedulability of this avionics system following the procedure in section \ref{sec:comp}:

(1) \emph{Decomposition:} The system is first decomposed into five sets of SWA template instances corresponding to five partitions. The schedulability of any partition $P_i,i=\{1,2,3,4,5\}$ is described as the \uppaal query $q_i$:
\begin{equation}
\uppPropAG{not perror[i]},
\end{equation}
where the boolean variable \uppVar{perror[i]} should be assigned to \uppConst{True} once any error locations are reached in $P_i$. When analyzing the schedulability of $P_i$, we \emph{only instantiate} the set of SWA template instances of $P_i$ into \uppaal processes. This set contains two scheduler models coming from \uppTemp{PartitionSupply} and \uppTemp{TaskScheduler}, all the \uppTemp{PeriodicTask} and \uppTemp{SporadicTask} models in $P_i$, and the communication layer models from which $P_i$ receives messages.

(2) \emph{Construction of message interfaces:} The message interfaces are constructed from the template depicted in Fig.\ref{fig:mieg}, for all the messages originate in periodic tasks. There are four unknown parameters \uppConst{period}, \uppConst{initOffset}, \uppConst{offset}, and \uppConst{jitter} in the template. Initially, the parameters of a message interface are set to the same values as these of the source task. Then we employ a binary search to heuristically refine \uppConst{offset} and \uppConst{jitter}, meanwhile guaranteeing timed selection simulation relation exists.

(3) \emph{Model checking:} The schedulability of five partitions is checked individually. After combining the models of $P_i$ and its message interfaces, we verify the property $q_i$ by model checking in \uppaal. The verification was repeated for each partition to evaluate the schedulability of a complete system. The experiment was executed on the \uppaal 4.1.19 64-bit version and an Intel Core i7-5600U laptop processor.

(4) \emph{Deduction:} According to the assume-guarantee rule described in Eq.(\ref{eq:deductionrule}), we conclude the schedulability of the complete system from the results of the verification of five partitions.

\vspace{-10pt}
\subsubsection*{Results of the Analysis}

The result in Table \ref{tab:timemem} shows that each partition is separately schedulable (The results ``Yes'' of Case 1) except the partition $P_3$ (The result ``No''). From a global view, we cannot conclude directly that the system is non-schedulable, because the compositional approach described in section \ref{sec:comp} only provides a sufficient condition for schedulability. Nevertheless, we find a counter-example by simulation in \uppaal, and thus it can be concluded that the current system is not schedulable. The counter-example shows that $P_3$ violates the constraint of the refresh period of $\mathit{Msg_2}$ due to network latency.

Considering the effect of network latency on the scheduling configuration, we updated the partition schedules by performing a swap of time slots between $P_1$ and $P_2$. The modified partition schedules provide five windows $\langle P_1,5,5\rangle$, $\langle P_2,0,5\rangle$, $\langle P_3,10,5\rangle$, $\langle P_4,15,5\rangle$, and $\langle P_5,20,5\rangle$. The compositional analysis of the updated system was executed again. The result (Case 2 in Table \ref{tab:timemem}) shows that all the partitions of the updated system are individually schedulable. Thus, the updated system finally achieves the schedulability at the global level.

Table \ref{tab:timemem} also shows the performance in terms of execution time and memory usage. In both cases, the partition $P_3$ contains more instantiated models (19 processes) than the other four partitions. As a result, model-checking runs evidently slower and requires more memory than the others. Nevertheless, the compositional analysis could be performed on ordinary computers within an acceptable time.

Compared with the compositional way, global analysis based on the same \uppaal models would require 51 processes including all the 22 task models, whose state space is much more complex than the others. This causes \uppaal to run out of memory within a few minutes, and thus makes the global analysis infeasible. In contrast, the compositional approach only requires at most 5 task models when we perform model checking, offering effective state space reduction.

\section{Conclusion}\label{sec:conclusion}

In this paper, we present a compositional approach for schedulability analysis of DIMA systems, which are modeled as a set of stopwatch automata in \uppaal, describing schedulability as safety properties of models. We check each ARINC-653 partition including its communication environment individually, thereby reducing the complexity of model-checking. The techniques presented in this paper are applicable to the design of DIMA scheduling systems. We have applied the compositional approach to a concrete DIMA system. As future work, we plan to develop a model-based approach to the automatic optimization and generation of the partition schedules of a DIMA system.
\begin{table}[!t]
\centering
\caption{The Experiment Results (Result), Execution Time (Time/sec.) and Memory Usage (Mem/MB)}
\label{tab:timemem}
\begin{tabular}{c|c|r|r|c|r|r}
\hline\hline
\multirow{2}{*}{No.}      &\multicolumn{3}{c|}{Case 1}                            &\multicolumn{3}{c}{Case 2} \\\cline{2-7}
                                &Result&Time&Mem           &Result&Time&Mem   \\
\hline
$P_1$                           &Yes&7.46&146                    &Yes&6.07&105            \\
$P_2$                           &Yes&0.95&46                     &Yes&1.10&52             \\
$P_3$                           &No &42.94&664                   &Yes&256.48&3041         \\
$P_4$                           &Yes&0.69&43                     &Yes&0.68&43             \\
$P_5$                           &Yes&19.41&509                   &Yes&128.56&2041         \\
\hline\hline
\end{tabular}
\end{table}

\bibliographystyle{eptcs}
\bibliography{dima}

\end{document}